\documentclass[reprint,
 amsmath,amssymb,longbibliography,
 aps]{revtex4-2}
 
\usepackage{amsmath}
\usepackage{nicefrac,xfrac}
\usepackage{graphicx}
\usepackage{dcolumn}
\usepackage{bm}

\begin{document}

\title{Codes for entanglement-assisted classical communication}

\date{August 7, 2024}
 \author{Tushita Prasad}
 \altaffiliation{\href{mailto:tushita.prasad@phdstud.ug.edu.pl}{tushita.prasad@phdstud.ug.edu.pl}}
\author{Markus Grassl}%
 \email{markus.grassl@ug.edu.pl}
\affiliation{%
International Centre for Theory of Quantum Technologies, University of Gdańsk, Jana Bażyńskiego 1A, 80-309 Gdańsk
}

\begin{abstract}
Entanglement-assisted classical communication (EACC) aims to enhance
communication systems using entanglement as an additional resource.
However, there is a scarcity of explicit protocols designed for finite
transmission scenarios, which presents a challenge for real-world
implementation. In response we introduce a new EACC scheme capable of
correcting a fixed number of erasures/errors.  It can be adjusted to the
available amount of entanglement and sends classical information over
a quantum channel. We establish a general framework to accomplish
such a task by reducing it to a classical problem. Comparing with
specific bounds we identify optimal parameter ranges. The scheme
requires only the implementation of super-dense coding which has been
demonstrated successfully in experiments. Furthermore, our results
shows that an adaptable entanglement use confers a
communication advantage.  Overall, our work sheds light on how
entanglement can elevate various finite-length communication
protocols, opening new avenues for exploration in the field.
\end{abstract}

\maketitle

\section{Introduction}

\begin{figure*}
\includegraphics[width=2\columnwidth]{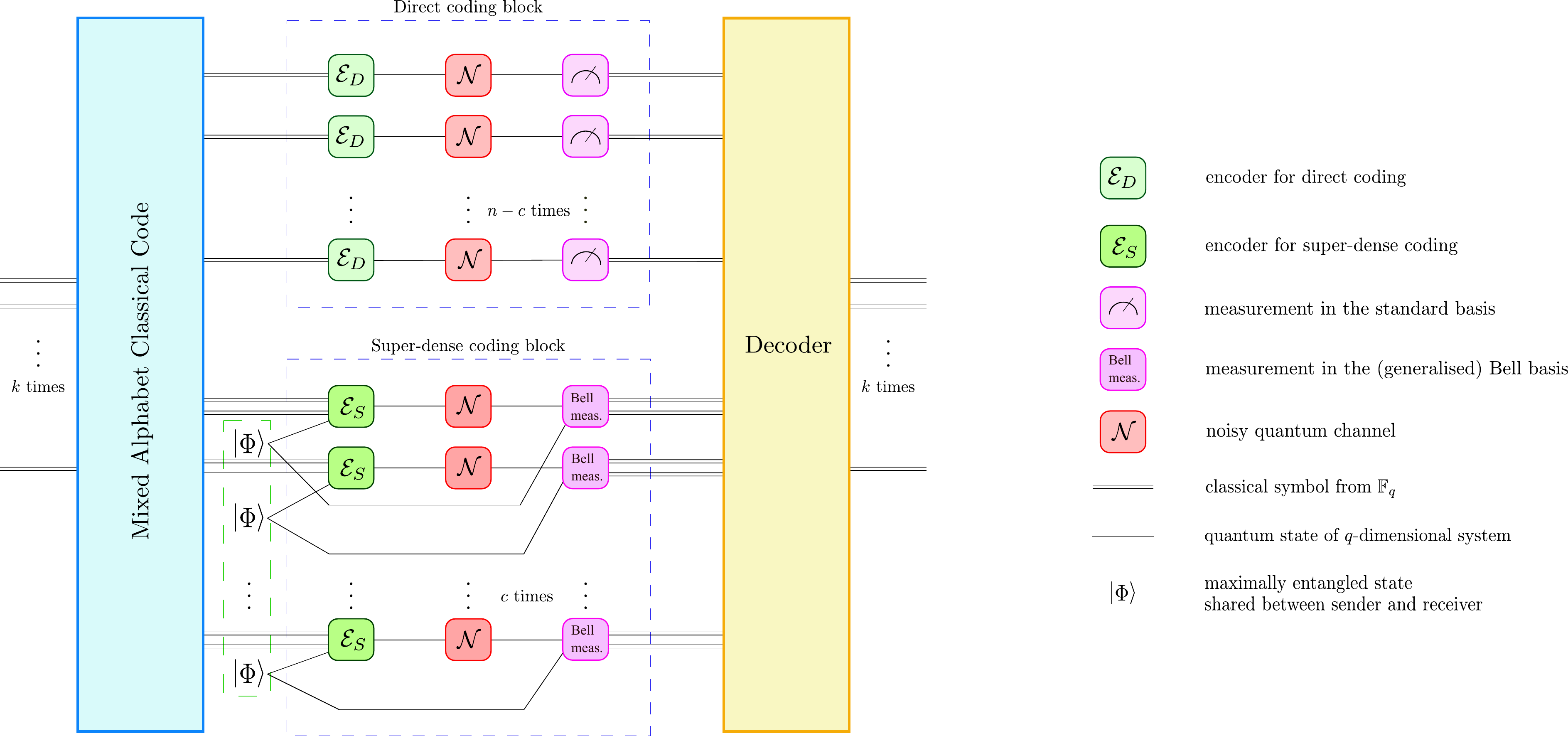}
\caption{\label{fig:scheme} \textbf{Schematic representation of the encoding and decoding process.} The mixed alphabet classical code maps $k$ input symbols to $n$ outputs from different alphabets. Then, $n-c$ of the symbols are transmitted using direct coding and $c$ symbols are transmitted using super-dense coding. The encoder $\mathcal{E}_{D}$ maps one classical symbol from $\mathbb{F}_q$  onto a qudit which is transmitted across a noisy quantum channel $\mathcal{N}$ and measured in the corresponding basis. This process is repeated $n-c$ times as illustrated by the \emph{direct coding block}. 
The encoder $\mathcal{E}_{S}$ encodes a pair of classical symbols (identifying $\mathbb{F}_{q^2}$ and $(\mathbb{F}_q)^2$) onto one half of a maximally entangled state $|\Phi\rangle$. This encoded state is then transmitted across a noisy quantum channel $\mathcal{N}$.  The other half of the maximally entangled state $|\Phi\rangle$  is shared with the receiver noiselessly. Upon reception, the receiver performs a measurement in the Bell basis on the received pair of states. This process is repeated $c$ times as illustrated by the \emph{super-dense coding block}. Finally, the decoder corrects errors and outputs a classical message.}
\end{figure*}

Entanglement serves as a valuable resource in many quantum communication and quantum information processing tasks. Pre-existing entanglement can sometimes enhance the information transmission capabilities of quantum channels. One of the earliest results demonstrating this power of entanglement-assisted communication came in 1992. Wiesner introduced the technique of super-dense coding \cite{wiesner} in which a sender can communicate two classical bits while sending a single qubit across a noiseless quantum channel, provided they have pre-shared entanglement. Subsequently, Bennett et al.~\cite{smolin} showed that the classical capacity of some noisy quantum channels can be significantly increased with the assistance of unbounded pre-shared entanglement between the sender and the receiver. In 2002, Bowen~\cite{Bowen} presented a class of entanglement-assisted (EA) quantum codes that achieve the quantum EA channel capacity when a certain amount of shared entanglement per channel use is provided. A few years later, Brun, Devetak and Hsieh \cite{devetak} proved that assisting a quantum channel with entanglement allows quantum codes to have better parameters than codes not aided by entanglement.

So far, entanglement-assisted communication has been largely explored through
an information theoretic lens. This means, prominently questions on
entanglement aided capacities have been addressed
\cite{bennett,shor,datta,hseih1,smolin, harrow, shadman,
  laurenza}. For instance, Hsieh and Wilde \cite{wilde} proved a
capacity theorem establishing the maximum rate at which one can
transmit classical and quantum information simultaneously across a
noisy quantum channel supported by entanglement. Shadman and Bruß
\cite{shadman}, on the other hand, consider a similar problem setting,
but focus on transmitting only classical information. They determine
the super-dense coding capacity for specific resource states in the
presence of noise. The interest in capacity questions is completely
natural considering how the concept of capacity forms a central theme
in information theory. Its significance can be seen in Shannon's
noisy-channel coding theorem \cite{shannon}, a foundational result
lying at the heart of information theory. 
Shannon showed that for a noisy channel, given an information
transmission rate below capacity, the transmission error probability
asymptotically tends to zero when the coding length increases.

The problem of reliable transmission of either classical or quantum information, or a combination of both, across $n$ uses of a quantum channel with the help of entanglement has been explored from a coding theory perspective as well, although this approach has been much less prevalent. This perspective involves addressing the problem in a practical setting, defined by specific parameters like a fixed code length and a set of errors which can be perfectly corrected. 
Kremsky et al.~\cite{kremsky} proposed a generalization of the stabilizer framework to capture entanglement-assisted hybrid codes. Here ``hybrid'' refers to the fact that these codes can transmit both classical and quantum information at the same time. The authors give examples of hybrid codes, but none that transmits only classical information. Their work primarily focuses on establishing a framework and conditions for error correction without addressing how to find codes that meet these conditions.
Also, they do not connect the parameters of the code described in their ``extended stabilizer formalism'' with the properties of the corresponding classical codes. In particular, there are no constructions in the literature for codes designed for entanglement-assisted classical communication.

While considerable attention has been devoted to exploring EA quantum codes, the exploration of EA classical codes has been notably scarce. A common assumption is that an unlimited amount of entanglement can be used, but the question of less entanglement has, for example, been considered in Ref. \cite{Bowen}. Directly using EA quantum codes to transmit classical information is inferior to the solution we present here. This is related to the fact that the parameters $[\![n,k,d;c]\!]_q$ of an EA quantum code obey the classical Singleton bound $d\le n-k+1$ given in eq.~\eqref{eq:SingletonBound}  below \cite{GrasslHuberWinter}, while our codes can exceed that bound.

Addressing these aspects, in this paper  we present an explicit encoding scheme that can transmit classical information using a qudit channel $n$ times assisted by $c\le n$ maximally entangled qudit pairs when at most $d-1$ erasures occur. We demand perfect error correction, i.e., the probability of error after decoding should be zero. 
Our scheme is ``explicit'' in the sense that we provide a construction method that works for all parameters within a certain range and allows to easily deduce the parameters of the resulting code.

The paper is structured as follows. In Sections \ref{sec:reduction} and \ref{sec:code_construction}, we address the task of transmitting classical information over a quantum channel using maximally entangled pairs. We simplify the problem by reducing it to a classical one and present an explicit encoding scheme, offering practical methods for creating directly implementable codes. In Section \ref{sec:bounds}, we derive Singleton-like bounds, compare our scheme to these bounds, and identify parameter ranges where the scheme is optimal. In Section \ref{sec:discussion}, we analyze how our proposed EACC scheme surpasses classical protocols in terms of error correction capability and information transfer rate. Finally, in Section \ref{sec:conclusion}, we conclude by discussing potential future directions and implications of our work.

\section{Results}
\subsection{ Reduction to a classical problem}\label{sec:reduction}
Let $q$ be the dimension of the quantum
system associated with the quantum channel.
We distinguish two cases. If we do not have entanglement available, for each use of the
quantum channel we can send $q$ mutually orthogonal states over the channel (one at a
time), and the receiver measures in the corresponding basis. In case we have
entanglement available, we can send $q^2$ different messages per use
of the channel with the super-dense coding protocol \cite{werner}
which requires one maximally entangled pair for each single
transmission. This allows for our problem to be translated to a
classical one. The amalgamation of direct coding and super-dense
coding results in a combination of classical channels: $n-c$ classical
channels transmitting one symbol each and $c$ classical channels
transmitting two symbols each. Each of the $c$ symbols being
communicated via the super dense-coding protocol can be interpreted as
two symbols from an alphabet containing $q$ letters.
The problem now reduces to finding an encoding of messages into a
string of $n$ symbols, where $n-c$ of them come from an alphabet
$\mathcal{A}_{q}$ with $q$ letters and $c$ symbols are from the alphabet
$\mathcal{A}_{q^2}=\mathcal{A}_q\times\mathcal{A}_q$ with $q^2$ letters.
An erasure during a single
transmission in the context of a super-dense coding scheme implies
that two classical symbols are erased, otherwise a single symbol is
considered to be erased. The setting is illustrated in Fig.~\ref{fig:scheme}.
In the extremal case of $c=0$ we can use a classical error correcting
code over the alphabet $\mathcal{A}_{q}$. When $n=c$ we can use a code
over $\mathcal{A}_{q^2}$. The case of $n=c$ has been considered earlier in Ref. \cite[Proposition 11]{brun}.
We denote the code by $\mathcal{C}=[n,k,d; c]_{q}$, where $n$ is the number of channel uses and $c$ is the amount of entanglement required by the code. The parameter $d$ is the minimum distance of the code. A code with minimum distance $d$ can correct a maximum of $d-1$
erasures or $\lfloor(d-1)/2\rfloor$ errors. An erasure is an error at a known position \cite{erasure}.  Further, $k$ is the dimension of the code, that is, $q^k$ different classical messages can be sent
across the channel. From here on, we assume $q$ to be a power of prime. Lastly, the alphabets $\mathcal{A}_q=\mathbb{F}_{q}$ and $\mathcal{A}_{q^2}=\mathbb{F}_{q^2}$ are chosen to be finite fields with $q$ elements and $q^2$ elements, respectively. 
For an introduction to finite fields and basics of coding theory see, for example, Ref. \cite{roman} or Ref. \cite{sloane}.

\subsection{Code construction}\label{sec:code_construction}
We use the following  modified version of the concept of Reed-Solomon codes \cite{reed} to construct the code. Let $\mathbb{F}_{q}[x]$ denote the set of polynomials in $x$ with coefficients in $\mathbb{F}_{q}$ and let $\mathbb{F}_{q}^{n-c}\times \mathbb{F}_{q^{2}}^{c}$ be the direct product of vector spaces.
We obtain our \textit{mixed alphabet Reed-Solomon code}  $[n,k,d;c]_{q}\subseteq \mathbb{F}_{q}^{n-c}\times \mathbb{F}_{q^{2}}^{c}$ by evaluating all polynomials $f\in \mathbb{F}_{q}[x]$ of degree at most $k-1$. More precisely,
\begin{multline}
\mathcal{C}=\Bigl\{\big(f(\alpha_{1}),\ldots ,f(\alpha_{n-c}),f(\gamma_{1}),\ldots,f(\gamma_{c})\big)\colon\\ f \in \mathbb{F}_{q}[x] ; \text{deg}\hspace{1mm}  f \leq k-1 \Bigr\}
\end{multline}
where $n-c = n_{1}$ distinct evaluation points from $\mathbb{F}_{q}$ and $c = n_{2}$
distinct points from $\mathbb{F}_{q^{2}}\setminus\mathbb{F}_{q}$ are
denoted by $\alpha_{i}$ and $\gamma_{j}$, respectively. The elements
of $\mathbb{F}_{q^{2}}\setminus\mathbb{F}_{q}$ come in conjugate pairs
$\{ \gamma, \gamma^{q}\}$. We choose only a single element $\gamma$ from each conjugate pair to be present amongst our evaluation points. This gives rise to the
partitioning of the $n$ evaluation points as $n = n_{1}+n_{2}$,
where $n_1$ is the number of elements from $\mathbb{F}_q$, and $n_2$ is the number of elements from $\mathbb{F}_{q^2}$. 

In the standard classical setting without entanglement assistance one cannot achieve a rate $R = k/n$ higher than one, since the size of the code $q^k$ cannot exceed the number $q^n$ of different strings over an alphabet with $q$ symbols. In contrast, our scheme allows $k$ to be larger than $n$ (whenever $c$ is non-zero) while maintaining an injective encoding. To see this, consider the situation at the receiver's end. In order to retrieve the encoded information from a received codeword, the receiver uses the fact that a polynomial of degree at most $k-1$ is uniquely determined by its value at $k$ evaluation points. For evaluation points $\gamma\in\mathbb{F}_{q^2}\setminus\mathbb{F}_{q}$,
we have $f(\gamma^q)=f(\gamma)^q$. Hence if the receiver knows the component of a
codeword corresponding to an evaluation point $\gamma_j$, he can also compute the evaluation at the conjugate point $\gamma_j^{q}$. This means that in the error-free case the receiver actually has $n_{1}+2n_{2}=n+n_2$ evaluation points at his disposal to identify the polynomial, i.e., interpolation is possible as long as the degree of the polynomial is strictly smaller than the number of evaluation points. Therefore we have $\deg(f)<n_{1}+2n_{2}$, or equivalently, $k-1<n_{1}+2n_{2}=n+n_2$.

The length of our code is limited by the maximum number of elements one can pick from $\mathbb{F}_{q}$ and the maximum number of single elements one can pick from $\mathbb{F}_{q^2}$, therefore we have $1\leq n\leq q+ (q^2-q)/2=(q^2+q)/2$.  We note that we can extend the length by one and increase
the minimum distance by one as well, using ``the point at infinity''. For simplicity, we do not consider that option in what follows.

For a linear code, the minimum distance $d$ of the code equals the minimum
weight (number of non-zero components) of a non-zero codeword. It can
be deduced from the maximum number of evaluation points that are roots
of a polynomial $f \in \mathbb{F}_{q}[x]$ of degree at most $k-1$.

For $n_{1} \geq  k-1$ we have
\begin{equation}
    d=n-k+1,\label{eq:Singleton}
\end{equation}
since there exists a polynomial $f$ of degree at most $k-1$ with coefficients in
$\mathbb{F}_{q}$ that has the maximal possible number of $k-1$ roots
among the $n$ evaluation points.
The distance in \eqref{eq:Singleton} meets the so-called (classical) Singleton bound
\begin{equation}\label{eq:SingletonBound}
d\le n-k+1
\end{equation}
with equality. The bound is derived as follows. A code with distance $d$ can correct $d-1$ erasures. Erasing $d-1$ of the $n$ symbols results in strings of length $n-d+1$. Since all these strings have to be different, we get the bound $k\le n-d+1$, which is equivalent to \eqref{eq:SingletonBound}.

For $n_{1} < k-1$, we have 
\begin{equation}
\begin{split}
    d &{}= n-\left(n_{1}+\lfloor(k-1-n_{1})/2\rfloor \right) \\
    &{}= \lceil (n-k+1+n_{2})/{2} \rceil\\
    &{}> n-k+1.
    \end{split}\label{eq:d_min_larger}
\end{equation} 
This is because of the following reasoning. We know that a polynomial of degree at most $k-1$ has at most $k-1$ roots. In the worst case scenario all of the $n_{1}$ evaluation points from the base field  $\mathbb{F}_{q}$ are roots, and the remaining $k-1-n_{1}$ roots come from the extension field  $\mathbb{F}_{q^{2}}$. If an element $\gamma$ in $\mathbb{F}_{q^2}$ is a root, its conjugate partner $\gamma ^{q}$ is also a root. Since we have only single elements present among our evaluation points, not their conjugate partners, this implies that, in the worst case, at most half of the remaining $k-1-n_{1}$ roots are among the $n_{2}$ elements. The inequality follows from the fact that $n_1<k-1$ implies that $n_2=n-n_1 >n-k+1$. Hence, the minimum distance in \eqref{eq:d_min_larger} exceeds the classical Singleton bound in \eqref{eq:SingletonBound}.

In the case of $n_{1} < k-1$ we have the freedom to choose $n_{1}$, $n_{2}$ according to $n=n_{1}+n_{2}$ and the constraints listed below such that the minimum distance \eqref{eq:d_min_larger} of a code $\mathcal{C}=[n, k, d ; c]_{q}$ is maximized. 
This optimization involves picking as many evaluation points $c=n_{2}$ from $\mathbb{F}_{q^{2}}\backslash \mathbb{F}_{q}$ as possible, constrained by $c={n_{2}}\leq (q^2 - q)/2$.  The number of evaluation points one can pick from $\mathbb{F}_{q}$ is restricted by the values of $n$, $c$ and $q$, giving us $n-c \leq n_{1}\leq q$. Also, we have $0\leq c\leq n$.

\subsection{Bounds}\label{sec:bounds}
We now examine different ranges of $n$, $c$ and $k$ to determine when our code is optimal with
respect to certain bounds.
\subsubsection{Block error bound} The size of our code, i.e., the number of classical messages that our code can transmit, can be bounded, assuming that the code can correct a maximum of $d-1$ erasures.

Each of the $n_{1}$ symbols in a codeword can be visualized as a block
of length one.  Each of the $n_{2}$ symbols in a codeword can be
seen as a block of length two. Overall, a codeword of length $n$
contains $n_{1}$ blocks from $\mathbb{F}_{q}$ and $n_{2}$ blocks
from $\mathbb{F}_{q^2}$. The blocks from $\mathbb{F}_{q^2}$ contain
two symbols from $\mathbb{F}_{q}$ per block, allowing us to identify
each element of $\mathbb{F}_{q^{2}}$ with an element in
$(\mathbb{F}_{q})^{2}$. This means erasing an element from the
extension field corresponds to erasing a block of two symbols. First
assume that $n_{2}\geq d-1$. Then, in the worst case all the
$d-1$ erasures occur within the $n_{2}$ blocks from
$\mathbb{F}_{q^2}$. After the erasures have occurred, we are left with
strings of $n_{2}-(d-1)$ symbols from $\mathbb{F}_{q^2}$ and
$n_{1}$ symbols from $\mathbb{F}_{q}$. All these strings have to be
different. This results in the bound
\begin{align}
    |\mathcal{C}|=q^{k} &{}\leq q^{2(n_{2}-(d-1))} \hspace{0.5mm} q^{n_{1}} \nonumber\\
  k &{}\leq 2 (n_{2}-(d-1))+n_{1}.\label{eq:block_error}
  \end{align}
 This is the block error bound. Essentially, in the analysis above we
 have customized the bound in Ref. \cite{block} to our situation. The
 block error bound \eqref{eq:block_error} is met when $n\leq q+ \frac{q^{2}-q}{2}$,
 $n-q\leq c$. In this case, the minimum distance is $d = \lceil(n-k+1+c)/2\rceil$
 as given in \eqref{eq:d_min_larger}. In the scenario when
 $n_{2}< d-1$, the block error bound reduces to the Singleton
 bound $k\leq n-d+1$. This bound is met as long as $n_{1} \geq
 k-1$, see \eqref{eq:Singleton}.

\subsubsection{Quantum bound}
We continue examining general bounds on the parameters of our setting. In this context, Ref.~\cite{winter} establishes a bound on hybrid codes that can transmit classical and quantum information simultaneously with the assistance of entanglement. We apply the bound in \cite{winter} to our scenario. Since we are not transmitting quantum information, their parameter $Q=0$ in our case. From Theorem 8 in Ref.~\cite{winter} and the corresponding equations (54), 
 (55) and (56) therein, we obtain the inequalities:
\begin{align}  
  k &{}\leq (n-d+1)(1+t)\label{eq:ineq6} \\
  -c &{}\leq (n-2d+2)t\label{eq:ineq7}\\
  k-c &{}\leq (n-d+1)-t(d-1)\label{eq:ineq8}
 \end{align}
 Here $t$ is a parameter such that $t\in [0,1]$. Note that in Ref.~\cite{winter}, all logarithms use base $2$, while our parameters $k$ and $c$ are related to base $q$. Hence the range $[0,\log_{2} q]$ for the parameter $t$ in Ref.~\cite{winter} yields the range $[0,1]$ in our setting. Also, the amount of classical information $k$ we are sending is $C$ in their case, and the amount of available entanglement $c$ for us is $E$ in their scenario. 

Consider the inequalities \eqref{eq:ineq6} and \eqref{eq:ineq8}. Fixing all but $k$ and $t$, we see that the maximum value of $k$ is achieved when the corresponding bounds on $k$ are equal. We compute the intersection point 
as $(n-d+1)(1+t)= c + (n-d+1)-t(d-1)$ giving us $t= c/n$. Particularly, we find that $t=c/n$ lies in the interval $[0,1]$. When we substitute $t=c/n$ in \eqref{eq:ineq7} we find that it is fulfilled. This means  $k_{\text{max}} = (n-d+1)(1+c/n)$.

Using the above analysis, we obtain the following
quantum bound
\begin{equation}
    k \leq (n-d+1)(1+c/n).\label{eq:quantum_bound}
\end{equation} 
Equivalently, this gives the following bound on the distance
\begin{equation}
    d \leq n+1-\frac{kn}{n+c}.\label{eq:quantum_bound_d}
\end{equation}
The main question that we seek to answer in this section is for which range of $n$, $k$ and $c$ the code constructed in Section \ref{sec:code_construction} is optimal. To reach the maximal integral value of $k$ in \eqref{eq:quantum_bound}, we get the condition
\begin{equation}
\Gamma = (n-d+1)(1+c/n)-k < 1.
\end{equation}
In this case we say that our code achieves dimension optimality.

To reach the maximal integral value of $d$ in \eqref{eq:quantum_bound_d}, we get the condition
\begin{equation}
\Delta = n+1 -\frac{kn}{n+c}-d < 1.
\end{equation}
In this scenario we say that the code achieves distance
optimality. Distance and dimension optimality are related as
follows. When $\Delta < n/(n+c)$, we have dimension optimality. When
$n/(n+c)\leq \Delta <1$, we have distance optimality, but not
dimension optimality, since fixing all other parameters of the code we
could increase the dimension by one without violating the bound
\eqref{eq:quantum_bound_d}. This means that dimension optimality is
harder to achieve and implies distance optimality as well.

For the minimum distance in \eqref{eq:d_min_larger} we have
$\Delta = n+1 -\frac{kn}{n+c}-\lceil (n-k+1+c)/2\rceil$. As a reminder, the minimum distance given in \eqref{eq:d_min_larger} is obtained after
imposing the conditions $n\le q+ (q^2-q)/2$ and $n-q \leq c$.
We know that when $n=c$, we can use super-dense coding and obtain optimal solutions. Therefore we only consider the case $n>c$.
The ranges of
$k$ for which our code is distance optimal are as follows
\[
    k>
\begin{cases}
 n+c -2\left (\frac{n+c}{n-c}\right )    ,& \text{if } n-k+1+c \hspace{1mm} \text{is odd;}\\
   \frac{(n+c)(n-c-1)}{n-c},  & \text{otherwise.}
\end{cases}
\]
    
When $n-k+1+c$ is even, we reach dimension optimality only for $k=n+c$ which implies $d=1$. However, when $n-k+1+c$ is odd, we reach dimension optimality when $k>n+c -\frac{2n}{n-c}$.

Note that for linear codes, the dimension $k$ is an integer, while the upper
bound \eqref{eq:quantum_bound} on the dimension is in general non-integral. This means that when the bound is non-integral, it might be possible to find larger codes, but they cannot be linear. However, as we have shown, in specific instances where the upper bound is integral, linear codes attain the bound \eqref{eq:quantum_bound}.

\section{Discussion}\label{sec:discussion}
\subsection{Quantum advantage: Surpassing Classical Codes using Entanglement}
In our protocol, entanglement enhances error correction by increasing the minimum distance of the codes. We can see this using
\begin{align*}
d=\lceil(n-k+1+n_2)/2\rceil
\end{align*}%
from eq.~\eqref{eq:d_min_larger} where the parameter $n_2$ equals the required amount of entanglement $c\ge n-k+1$.  If we fix $k$, with more entanglement we can achieve a higher minimum distance. At the same time, fixing the distance $d$, with more entanglement we get a higher dimension $k$ and in turn a higher rate. Therefore, in our scenario entanglement helps us both to increase the distance or to increase the rate.

Moreover, entanglement allows us to achieve significantly higher minimum distances than what is possible classically without entanglement. Consider the following example. A code $[34,20,22;28]_{8}$ using $c=n_{2}=28$ maximally entangled pairs achieves a minimum distance of $22$. In contrast, without the use of entanglement a linear code with parameters $[34,20,d]_8$ can attain a maximum minimum distance of only $12$. Furthermore, only a code $[38,20,10]_8$ is explicitly known \cite{Grassl:codetables}. To illustrate further, consider fixing the distance instead of the dimension. That is, we might ask what the largest possible code with distance $22$ could be. 
Without the use of entanglement, a linear code with parameters $[34,k,22]_8$ can achieve a maximum dimension of only 
$10$ \cite{Grassl:codetables}. In contrast, with entanglement assistance, we achieve twice the dimension, specifically $k=20$.

In general, entanglement helps to achieve better code parameters in
 our scheme. Consider the code rate $R=k/n$. As discussed earlier in this section, there exists a trade-off between maximizing the rate for efficient information transmission and increasing the minimum distance for reliable error correction. Finding the best trade-off is an important question in coding theory.

\begin{figure}
    \begin{minipage}[t]{0.44\textwidth}
        \raggedright
        \includegraphics[width=\hsize]{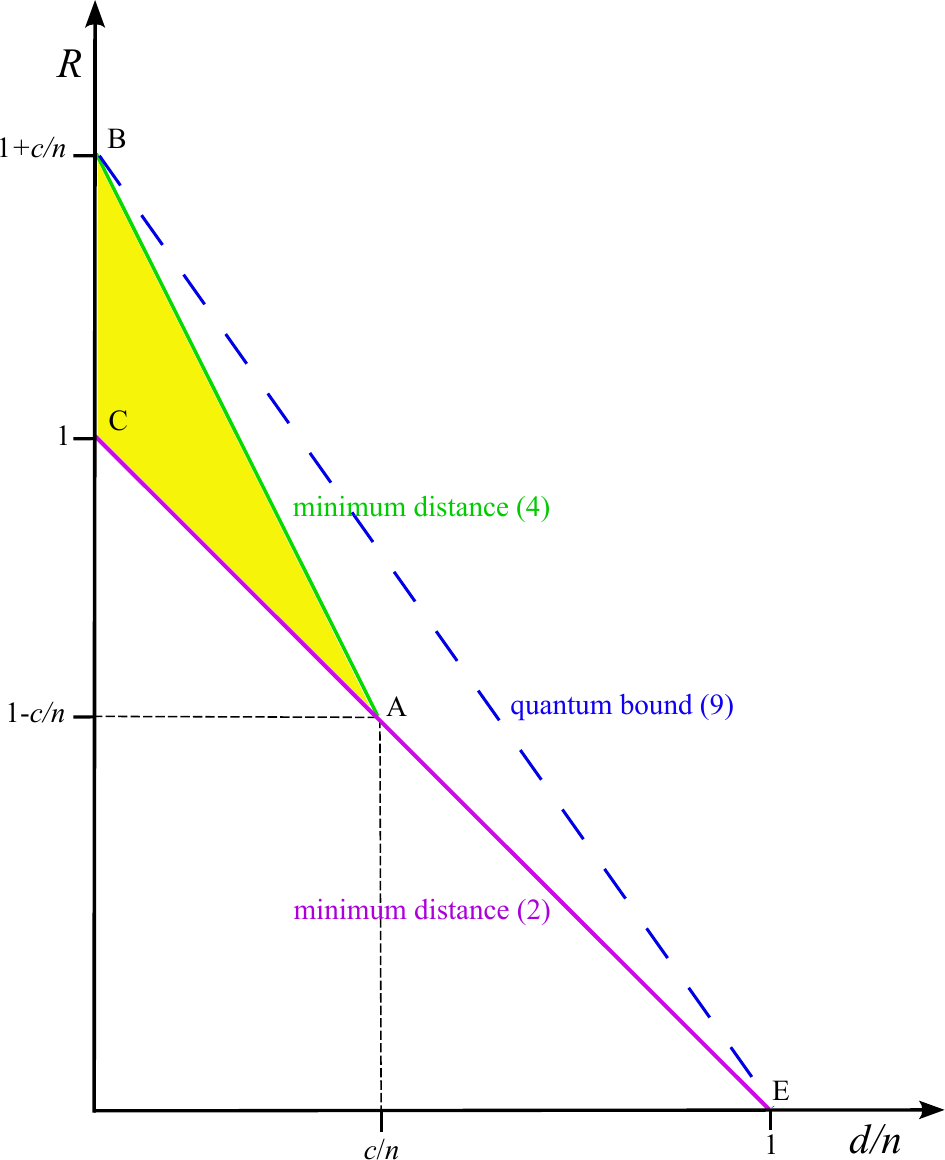}
        \caption{The code rate $R=k/n$ of our mixed-alphabet Reed-Solomon code is plotted as a function of the normalised minimum distance $d/n$ in the asymptotic limit.}
        \label{fig:graph}
    \end{minipage}
\end{figure}

In Fig.~\ref{fig:graph} we illustrate this trade-off  for the code we constructed in Section \ref{sec:code_construction}. We plot the code rate $R$ as a function of the normalised minimum distance $d/n$ in the asymptotic limit, ignoring additive terms that vanish when $n$ goes to infinity. When $n_{1}\geq k-1$ our code with minimum distance \eqref{eq:Singleton} achieves rates represented by the line AE. Conversely, when $n_{1}<k-1$, our code with minimum distance \eqref{eq:d_min_larger} achieves rates represented by the line BA. Note that the line segments BA and AE agree with the block error bound. The region ABC encompasses points where the rate is greater than one and also includes points with rates less than one, all of which still exceed the Singleton bound, indicating that we are surpassing the performance of any classical code. Entanglement is the reason we can reach these higher rates unattainable by classical means within this range. The line AE indicates the range when our codes meet the Singleton bound. While the length of a classical MDS code with $1<k<n-1$ cannot exceed $2q-2$ (see \cite[p. 94]{ball}), our entanglement-assisted codes can have length up to $(q^2+q)/2$.

\subsection{Other aspects}

 Our scheme allows to use the mixed alphabet
 Reed-Solomon codes introduced here. Using not only evaluation points
 from the field $\mathbb{F}_q$, but also from the field
 $\mathbb{F}_{q^2}$ we obtain longer codes, extending the length of
 optimal codes for fixed parameter $q$ from $O(q)$ to $O(q^2)$. The
 scheme is not restricted to these codes.  One can use other mixed
 alphabet codes or error-block correcting codes \cite{blockerror}.
 
 Also, our scheme is not restricted to the setting of perfect
 entanglement. Noisy entanglement (studied widely in the literature)
 results in errors in the super-dense coding protocol, which can
 equivalently be treated as errors during the transmission. 
 In the error correction process, 
 as long as the number of errors plus the
 number of ``bad'' entangled states is less than half of the minimum
 distance, our scheme will still work.

While our scheme provides an advantage over unassisted classical codes, we reach the quantum bound \eqref{eq:quantum_bound} only for specific parameters. We leave it to future research to determine whether there are entanglement-assisted codes for classical communication whose parameters lie in the region BAE of Fig. \ref{fig:graph}.

 Finally, we would like to point out that our scheme does not require
 quantum memory or quantum computation.  It only uses the
 primitive of super-dense coding that has been experimentally
 demonstrated \cite{super-dense_coding,super-dense_coding2}.

 \section{Conclusion.}\label{sec:conclusion}
 We present an explicit protocol together with its parameters for EA classical communication. Our research provides an avenue to illustrate how entanglement can boost various finite-length communication protocols. In certain scenarios, using entanglement can enable a code to correct twice as many errors compared to a code with similar parameters without entanglement assistance. Additionally, in specific cases, entanglement allows a code to achieve a rate higher than one, a feat which is not possible classically without entanglement. 

One of the open problems is to find coding schemes whose parameters are in the region BAE of Fig.~\ref{fig:graph}. As the classical codes underlying our scheme meet the block error bound, codes reaching the quantum bound BE are unlikely to be derived from our reduction to a classical problem.
 
Our work paves the way for finding new explicit code constructions as well. Also, hybrid protocols for simultaneously sending both classical and quantum information with the assistance of entanglement can potentially be obtained by combining our scheme with existing schemes for EA quantum communication. Developing such hybrid codes, facilitated by our code construction, has far reaching benefits. Explicit hybrid protocols have the potential to enhance error correction, secure communication and may play an important role in quantum computing tasks. 
\section{Acknowledgements} This research is supported by the `International Centre for Theory of Quantum Technologies' project
(contract no.~MAB/2018/5) carried out within the International
Research Agendas Programme of the Foundation for Polish Science
co-financed by the European Union from the funds of the Smart Growth
Operational Programme, axis IV: Increasing the research potential
(Measure 4.3).

\vfill


%

\end{document}